\documentclass[pra,twocolumn,aps,showpacs,10pt]{revtex4-1}

\usepackage{graphicx}  
\usepackage{dcolumn}  
\usepackage{amssymb, amsmath}
\usepackage{lineno} 
\usepackage{braket}
\usepackage{multirow}
\usepackage{bbold}
\usepackage{color}

\begin{document}
\title{Robust solid $^{129}$Xe longitudinal relaxation times}
\author{M.\ E.\ Limes}
\email{mlimes@princeton.edu}
\author{Z.\  L.\ Ma }
\author{E.\ G.\ Sorte}
\author{B.\ Saam}	
\email{saam@physics.utah.edu}

\affiliation{Department of Physics and Astronomy, University of Utah, 115 South 1400 East, Salt Lake City, Utah, 84112-0830, USA}

\begin{abstract}
We find that if solid xenon is formed from liquid xenon, denoted ``ice'', there is a 10\% increase of $^{129}$Xe longitudinal relaxation $T_1$ time (taken at 77 K and 2 Tesla) over a trickle-freeze formation, denoted ``snow''. 
Forming xenon ice also gives unprecedented reproducibility of $^{129}$Xe $T_1$ measurements across a range of 77-150 K.
This temperature dependence roughly follows the theory of spin-rotation mediated by Raman scattering of harmonic phonons (SRRS), though it results in a smaller-than-predicted spin-rotation coupling strength $c_{K0}/h$.
Enriched ice $^{129}$Xe $T_1$ experiments show no isotopic dependence in bulk relaxation mechanisms at 77 K and at kilogauss fields.  
\end{abstract}

\maketitle
%
\section{Introduction}

An effective and widely used methodology \cite{Driehuys_1996,Ruset_2006,Schrank_2009} for producing large quantities of hyperpolarized $^{129}$Xe includes cryogenic condensation, accumulation, and storage in the solid state of $^{129}$Xe gas polarized by spin-exchange optical pumping \cite{Walker_1997} in a large magnetic field (a few kilogauss or greater). 
Despite recent progress in improving gas-phase storage methods \cite{Berry-Pusey_2006,Anger_2008,Moller_2011}, almost all commercial and home-built xenon polarizers employ cryogenic storage in a flow-through system. (The recently developed ``open source" polarizer is a notable exception \cite{Goodson_2013}.) 
A cold trap using liquid nitrogen (LN$_2$) is typically used both to separate xenon from the other gases in the mixture that are needed to optimize spin-exchange optical pumping (SEOP) and to provide a compact storage vessel with a polarization lifetime $T_1 \approx 2.5$~h. 
A detailed understanding of $^{129}$Xe longitudinal relaxation in solid xenon is thus important for its use in a wide variety of fundamental studies and applications \cite{Bulatowicz_2013,Sorte_2011,Schnurr_2013}, including magnetic resonance imaging of the lung \cite{Mugler_1997,Driehuys_2009,Dregely_2011,Kaushik_2013}.
More generally, noble-gas solids are excellent model systems for first-principles calculation of nuclear relaxation mechanisms and rates. The spin-rotation interaction is the dominant intrinsic mechanism for $^{129}$Xe relaxation under almost all conditions of temperature, applied field, and xenon phases. 
(Ref. \cite{Saam_2015} provides a thorough review of $^{129}$Xe relaxation across the range of these variables.)
The corresponding Hamiltonian is given by $c_K(r) \mathbf{K} \cdot \mathbf{N}$, with coupling strength $c_K(r)$ between a nuclear spin $\mathbf{K}$ and the rotational angular momentum $\mathbf{N}$ of a pair of xenon atoms separated by a distance $r$ \cite{Torrey_1963}.
In the gas and liquid phases, the rotational angular momentum $\mathbf{N}$ of the pair is provided by a colliding pair of Xe atoms---transient and/or persistent (van der Waals) dimers \cite{Streever_1961,Hunt_1963,Chann_2002,Anger_2008}.
In the solid phase, the orbital angular momentum of a neighboring Xe pair comes from the phonon bath, which can be analyzed to determine the expected temperature dependence of the relaxation. 
Early studies of solid $^{129}$Xe by Happer and co-workers \cite{Cates_1990,Gatzke_1993,Fitzgerald_1999} concluded that the spin-rotation interaction mediated by the Raman-scattering of harmonic phonons (SRRS) is responsible for high-field ($\gtrsim 1$~kG) relaxation of solid $^{129}$Xe in the temperature range of 50-120 K. 
\begin{figure}[t!]
\center
\includegraphics{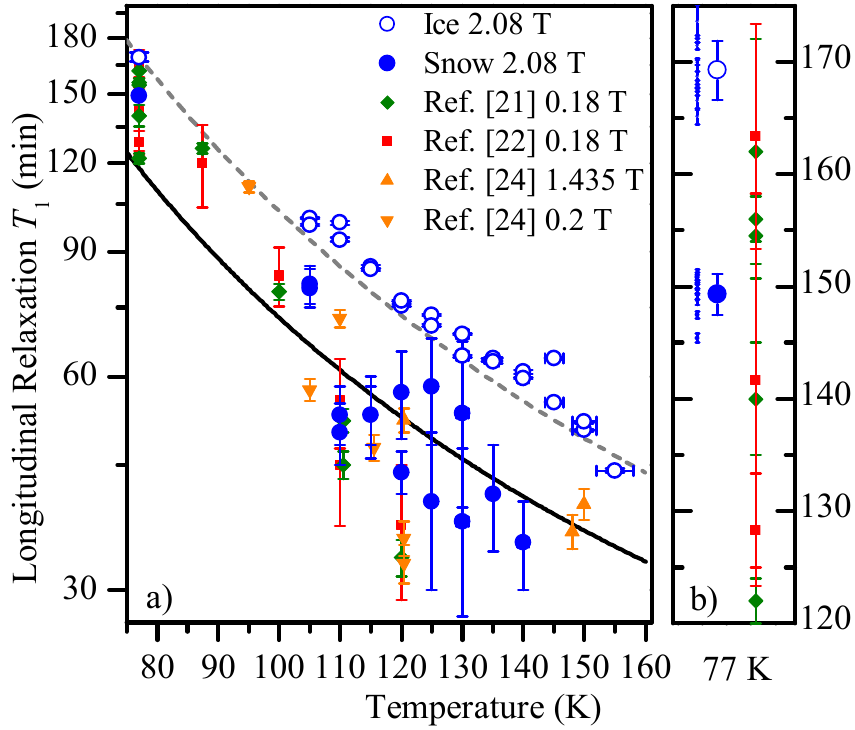}
\caption{ All $^{129}$Xe $T_1$ data from this work shown plotted vs.~temperature for both naturally abundant ice and snow, as well as comparable data from Refs.~\cite{Cates_1990, Gatzke_1993,Kuzma_2002} along with indicated applied magnetic field. 
The black line is the theory of spin-rotation relaxation mediated by Raman scattering (SRRS) from Ref.~\cite{Fitzgerald_1999} with $c_{K0}/h = -27$ Hz. 
The grey line is the result of forcing the SRRS through our robust $T_1$ value for ice at 77 K, yielding $c_{K0}/h = -22.3 \pm 0.2$ Hz and also shows significant discrepancy at higher temperatures.  
The large error bars for our higher temperature snow data correspond to the observed multi-exponential behavior. 
A detailed view of our ice and snow data at 77 K is shown in (b), where we plot the range of our individual measured values (left), weighted averages for both ice and snow (middle), and the more widely varying 77 K data from Refs.~\cite{Cates_1990} and \cite{Gatzke_1993} (right). 
The weighted averages show a $T_1$ in ice to be consistently $\approx$10\% longer than in snow. 
}
\label{fig:snow_ice_temp.pdf}
\end{figure}
Kuzma, {\em et al.}~\cite{Kuzma_2002}, later extended this work, particularly in noting the potency of the vacancy-diffusion mechanism \cite{Yen_1963} for low-fields and temperatures approaching the melting point (160~K), and the corresponding importance of thermal management of the accumulating solid in a practical polarizer (i.e., maintaining the entire sample at or near 77~K). 
The SRRS theory is an extension of that developed first by van Kranendonk and Walker \cite{vanKranendonk_1954,vanKranendonk_1967,vanKranendonk_1968} for the case of quadrupolar nuclei (spin $> 1/2$), for which the shorter $T_1$ values are experimentally more accessible in the absence of hyperpolarization. 
As such, it should be clear that SRRS is a {\em bulk} mechanism, i.e., the theory does not predict any dependence to the longitudinal relaxation on crystallite size, although it does depend on the relevant Bravais lattice (fcc for xenon) \cite{Fitzgerald_1999,Kuzma_2002_2,Vega_2006}.

In this work, we show a dependence of longitudinal relaxation ($T_1$) of $^{129}$Xe on freezing method, in particular a difference in our ``snow" and ``ice" $T_1$'s, and a deviation from the predicted spin-rotation coupling constant. 
A summary of the results of this paper, as well as a collection of relevant past results, is shown in Fig.~\ref{fig:snow_ice_temp.pdf}.
Our data at 77 K show a roughly $10$\% discrepancy in $T_1$ between a trickle-frozen sample, denoted snow, and a sample that has been taken through the liquid phase of xenon and re-frozen, denoted ice. 
Our temperature-dependent ice data show unprecedented reproducibility across a range of 77-150 K, and is approximately 30\% longer than the prediction from the SRRS theory in Ref.~\onlinecite{Fitzgerald_1999}. 
A negligible isotope effect for ice is also shown, which agrees with previous experiments \cite{Cates_1990}.

\section{Methods}
For most experiments, we used $^{129}$Xe polarized by a home-built, flow-through polarizer, for which detailed specifications may be found in Ref.~\onlinecite{Schrank_2009}. 
In brief, a gas mixture lean in naturally abundant xenon (Linde) was allowed to flow upward through a long vertically oriented cylindrical glass optical pumping cell (1 m long by 10~cm dia). 
The SEOP laser was a 50-W single-bar diode-laser array (Model M1B-795.2-50C-SS4.1E, DILAS) that was frequency-narrowed by an external Littrow cavity \cite{Chann_2000}.
The laser had integral collimating lenses for both fast and slow axes, and we used an improved optics scheme with a $\lambda/2$-plate and polarizing beam-splitter cube to establish a separate low-power narrowing channel \cite{Chann_2000,Buchta_2007}. 
We operated the polarizer at a cell pressure of about $1.1$~bar with a gas mixture that had partial flow rates in the ratio He:N$_2$:Xe = 1000:500:10 sccm. 
The gases were mixed and sent through a purifier (model FT400902, SAES PureGas) to remove oxygen and other impurities before entering at the bottom of the polarizing cell, where the Rb vapor also acted as a getter for oxygen. 
SEOP took place in the lower third of the cell, which sat inside of an oven held at 140$^{\circ}$C. 
In the experiments reported here, the Rb and $^{129}$Xe were polarized into  low-energy Zeeman states.

\begin{figure}
\includegraphics{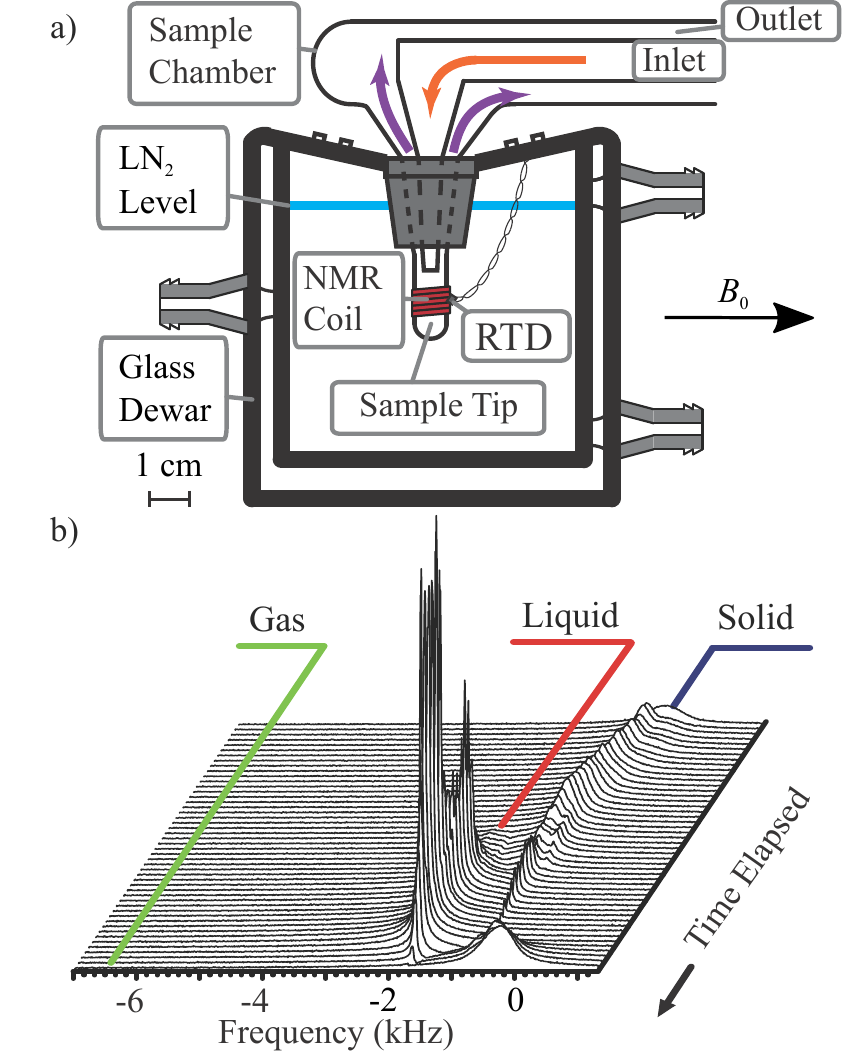}
\caption{ (a) A schematic of the sample chamber used to collect xenon and perform NMR measurements.
The top of the glass dewar was sealed with a rubber stopper through which a re-entrant glass-tube assembly is positioned. The central inlet tube was connected to the PTFE (Teflon) tubing that carries in the polarized xenon gas from the polarizer; the coaxial outlet tube vented to the atmosphere through a rough vacuum pump. Xenon snow was condensed from the polarized-gas stream and collected at the bottom of the outlet tube (sample tip). Cold nitrogen gas from a pressurized dewar (not shown) flowed through the sample chamber to provide measurement temperatures in the range of 105--155~K, plus measurements at 77~K when the chamber was flooded with LN$_2$. (b) A time sequence of $^{129}$Xe NMR spectra (signal intensity vs. relative frequency) indicates the positions of the gas, liquid, and solid peaks, as all phases have distinct chemical shifts. (Although the gas line is not visible here, we detected the shift in other experiments and obtained a value of $T_1 \approx 82$ min in the sample chamber.) The roughly $3$~min sequence shows the initial formation of snow, complete melting of the snow to liquid at 160--175~K, and refreezing of the sample at 77~K to make ice.}
\label{fig:schematic}
\end{figure}

After emerging from the top of the cell through a valve, the now-polarized gas mixture was directed through about $12$~m of 6-mm i.d.\ PTFE tubing, across a hallway and into a second nearby laboratory housing a 2~T horizontal-bore Oxford superconducting magnet. 
The tubing was connected directly to a sample chamber located at the magnet isocenter (a schematic is shown in Fig.~\ref{fig:schematic}a). 
The sample chamber was a small glass dewar through which cold nitrogen gas could flow, or in which LN$_2$ could be allowed to accumulate. 
A co-axial re-entrant tube assembly was positioned inside the chamber. 
The bottom of the outlet tube (sample tip) sat centrally in the dewar and was held at 77~K for initial collection of a snow sample.
Xenon was frozen out of the gas mixture as it flowed down through the middle and then back up and out along the sides of the sample chamber. 
A typical collection period was 20 min, producing roughly 10 mmol of hyperpolarized solid xenon. 

A solenoidal NMR probe coil, tuned to the $\omega_0 /2\pi = 24.6$~MHz Larmor frequency of $^{129}$Xe at 2.08~T, was wrapped around the sample tip, and a thin-film resistive thermal device (RTD; model F3105, Omega Engineering) was placed nearby.  A few of the initial measurements were conducted using an Apollo (Tecmag) NMR spectometer; the majority were made using the later-model Redstone (Tecmag). 
During collection of the xenon snow and throughout the subsequent experimental run, the NMR spectral intensity of the gas, liquid, and solid peaks, all of which have different chemical shifts, were monitored using low-flip-angle pulses (typical pulse length 0.5-5~$\mu$s resulting in polarization losses $<1 \%$) fed into a 250~mW rf amplifier (model ZFL-1000VH, Mini-Circuits). 
After collecting a snow sample at 77~K, there were two possible experimental protocols: (1) the snow sample was brought to measurement temperature, the relative polarization was recorded using a standard low-flip-angle pulse, and a $T_1$ measurement was made; or (2) the sample was completely liquefied by heating it to 160-170~K and then refrozen at 77~K to form ice, and a $T_1$ measurement was made. 
In the latter case, we monitored the NMR spectrum to ensure that the sample became entirely liquid prior to refreezing; see Fig.~\ref{fig:schematic}(b). 
The snow or ice sample was brought to the measurement temperature by flowing pressurized boil-off from a LN$_2$ dewar through the sample chamber. A temperature controller (model CNi16D33, Omega Engineering) monitored the RTD and switched a nichrome-wire heater positioned in the path of the flowing cold gas to control the temperature at a pre-determined set point between 105 and 155~K. For 77~K data points, the sample chamber was simply flooded with with LN$_2$.

\begin{figure}
\center
\includegraphics{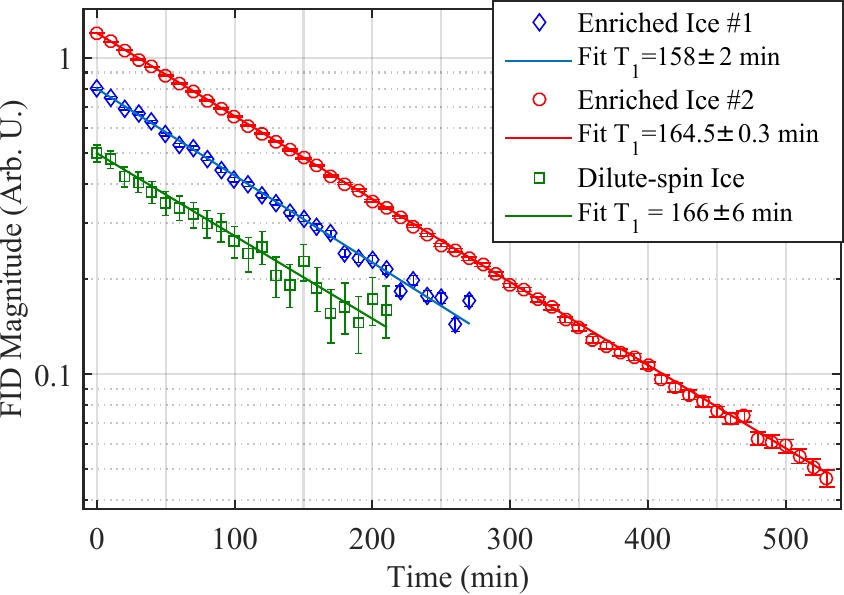}
\caption{Results from enriched 86\% $^{129}$Xe ice, and 4.5\% $^{129}$Xe dilute-spin ice, taken at 77 K and 2 T. As the concentration of $^{129}$Xe $N_K$ varies by a factor of 20 in these experiments, any bulk paramagnetic relaxation due to dilute impurities (which would be proportional to $N_K$) is ruled out for our ice samples. The second enriched ice sample was formed after an enriched snow experiment was ran.}
\label{fig:isotope_dependence}
\end{figure}

A series of additional experiments were conducted to test for any dependence of $T_1$ relaxation on the isotopic abundance of $^{129}$Xe; see Fig.~\ref{fig:isotope_dependence}. 
Some of these measurements, for which the $^{129}$Xe concentration was below natural abundance (``dilute-spin" experiments), were conducted with sealed convection cells, with liquid xenon polarized via phase exchange with its gas-phase vapor pressure of hyperpolarized xenon \cite{Su_2004,Morgan_2008,Sorte_2011}. 
These cells were made by mixing naturally abundant xenon with xenon that is isotopically enriched in spinless isotopes. 

The measurements conducted for this work are conceptually simple: we monitored the decay of the hyperpolarized magnetization of the sample vs.~time by acquiring periodic free-induction decays with low-flip-angle pulses.
NMR signals from solid hyperpolarized $^{129}$Xe are very large; the large dynamic range of signals led us to conduct careful linearity tests of the spectrometer's amplifier chain across the range of settings used. 
We also used both high-Q ``tank'' and low-Q ``flat'' probes,
and found no significant differences in the obtained data. 
For each $T_1$ measurement an appropriate pulse size was chosen such that the spectrometer's ADC buffer was as full as possible without saturating the receiving chain of amplifiers, and the flip angle could be calibrated.
The data were fit to:
$S(t) = S_0e^{-t/T_1},$
where the initial signal $S_0$ is a fit parameter, $T_1$ is the extracted relaxation time, and typically $\Delta t = 5$~min between each FID. 
After a $T_1$ measurement, the sample chamber was warmed and evacuated in preparation for the next run; the sealed convection cells were simply warmed and repumped.

Because our $T_1$ measurements are conducted at large polarizations, on the order of 10\%, we assumed for our analysis that $S(\infty) \approx 0$, i.e., that the signal from the sample in thermal equilibrium was negligible. 
However, the thermal signal was measurable, and we used it to obtain an estimate of the typical absolute $^{129}$Xe polarization in the initially formed snow sample by comparing its magnitude to that of the hyperpolarized signal. 
After acquiring a snow signal at 77~K, the sample magnetization was destroyed with rf pulses. 
The sample was then left at 77~K for 13~h (many times $T_1$), in order for the polarization to reach its thermal-equilibrium value, which we calculate as
$P_{\text{eq}} \approx \hbar\omega/2kT=7.6\times 10^{-6}.$
 Using our measured ratio of hyperpolarized to thermally polarized signals of $3.57\times 10^4$, we obtained a typical value of $P_{\text{hyp}} = 27\pm 5 \%$ for the initial polarization of the hyperpolarized snow, where the error is dominated by the roughly $20\%$ relative error in the determination of the weak thermal-equilibrium signal. 
 We note that during the thaw-refreeze process to make ice, we typically incurred a $10-20 \%$ fractional loss in polarization. 
The liquid phase $T_1$ is known to be roughly $25$~min~\cite{Sauer_1997, Savukov_2006}; liquid-phase relaxation is thus the likely cause of most of this loss and would limit the utility of converting snow samples to ice in order to improve storage times. 

\section{Results and Discussion} 

The results of naturally abundant ice and snow $^{129}$Xe $T_1$ measurements at 77~K are shown in Fig.~\ref{fig:snow_ice_temp.pdf}(b). 
The scatter in the data is small and averages to 169 $\pm$ 3 min for ice and 149 $\pm$ 2 min for snow. 
Although the $T_1$ difference between snow and ice is relatively small, the respective weighted-average values are robust and reproducible.
We varied the experimental arrangement in several ways, in an attempt to discount any methodological errors that could be present; these included using different probe designs, evacuating the buffer gases in the sample chamber prior to measurement, and using different methods for temperature control.
All these variations produced insignificant changes in the ice-snow discrepancy at 77 K.

The most obvious physical difference between snow and ice is the much greater surface-to-volume ratio for the snow. 
The snow appears white and should thus consist of crystallites significantly larger than optical wavelengths---1-100 $\mu$m, whereas the ice appears as a cylindrical translucent plug several mm in length and diameter. 
Due to our sudden-freeze technique, the ice is unlikely to consist of any appreciably-sized single crystals of xenon \cite{Keyse_1985, Maruyama_1988}. 
We conclude only that since the ice does not scatter as much light, it has fewer grain boundaries and a smaller overall surface area than the snow. This leads directly to the hypothesis that the snow-ice discrepancy is due to an additional relaxation mechanism acting only at the suface of snow crystallites, with further relaxation then proceeding through diffusion of magnetization in the bulk to the surface.
One such mechanism is believed responsible for the residual high-field field dependence of $T_1$ in snow at 4~K, observed by Gatzke, {\em et al.} \cite{Gatzke_1993}.
In that work, the authors propose thermal mixing of $^{129}$Xe and extreme quadrupolar-broadened  $^{131}$Xe nuclei (21\% abundant in natural xenon) near crystallite surfaces (where there are substantial electric-field gradients). 
We can safely discount this mechanism in our work at 77~K and above because it only contributed significantly in Ref.~\onlinecite{Gatzke_1993} at relaxation times measuring many hundreds of hours, as well as the mechanism being additionally suppressed at high magnetic fields---our $H_0\approx 2$~T field is an order of magnitude larger than that used in Ref.~\onlinecite{Gatzke_1993}. 

A more likely possibility for diffusion-driven relaxation is the adsorption of paramagnetic oxygen onto crystallite surfaces.
Flow-through polarizers and similar designs that involve cryogenic separation of xenon (and later revolatilization outside of the optical-pumping cell) open the possibility that residual oxygen may be present with the frozen solid; by contrast, earlier studies, such as those in Refs.~\onlinecite{Cates_1990, Gatzke_1993}, froze the xenon inside the same sealed cell in which it was polarized, meaning that any residual oxygen was likely gettered by the rubidium metal needed for SEOP. 

A hopping rate of roughly $0.2$~h$^{-1}$ per xenon atom at 77~K can be obtained \cite{Kuzma_2002} from the temperature-dependent transverse-relaxation ($T_2$) data reported by Yen and Norberg \cite{Yen_1963}.
 Given that the hopping distance is a few angstroms, atomic diffusion across distances on the order of 1~$\mu$m would take millions of hours.  
Dipolar spin diffusion is much faster: $^{129}$Xe in a naturally abundant, rigid, zero-temperature lattice has an estimated spin-diffusion coefficient of $D = 7\times 10^{-14}$~cm$^2$/s \cite{Gatzke_1993,Gatzke_1992}.
Considering spin-diffusion to a fast-relaxing boundary and the slowest decaying mode of a spherical grain, we crudely estimate an average crystallite radius required to account for the ice/snow difference as $ R = \sqrt{\pi^2 D T_1'} \approx 2 \mu$m, with $1/T_1'=(1/T_{1 Snow} - 1/T_{1 Ice})$. 
This estimate is within reason, so we assume spin-diffusion to a fast-relaxing boundary is the most likely candidate for the observed ice/snow discrepancy. 
However, as discussed below, the exact cause of the increased boundary relaxation, whether it be oxygen or a basic effect of grain boundaries, has yet to be determined. 

A key prediction of the SRRS theory is its temperature dependence of the relaxation time, thus we made $T_1$ measurements on separate ice samples across the range 105~K to 155~K in 5~K increments (temperatures were controlled to $\pm 0.5$~K up to 135~K, $\pm 1$~K for the 140~K and 145~K, $\pm 2$~K for the 150~K, and $\pm 3$~K for 155~K). 
All ice data were strictly mono-exponential, consistent with only uniform bulk relaxation mechanisms at the temperatures examined. 
Ice $T_1$ values ---are reliably longer than those for snow across this temperature range; indeed they appear to represent an upper limit with respect to all previously published measurements.
The notable exception to this is the highest-temperature data point at 155~K, for which the relaxation rate falls below the SRRS curve. 
This may be due to the onset of vacancy-diffusion processes at higher temperatures; even in a 2~T magnetic field, the high-end tail of the hopping-frequency distribution can still contribute significantly to the relaxation  \cite{Kuzma_2002}.

The harmonic two-phonon SRRS theory gives the relaxation rate for a single spin-1/2 nucleus due to coupling to the rotational momentum of each of its twelve nearest neighbors.
Our ice $T_1$ values are roughly 30$\%$ longer than predicted by the SRRS theory across the our observed temperature range. 
While observing $T_1$ values shorter than predicted may always be attributable to some sample impurities or additional overlooked mechanisms, consistently observing values longer than predicted require some re-examination of the theory. 
We do not repeat the details of the SRRS theory here, but wish to highlight that the prediction of the dependence of $T_1$ on temperature $T$ and spin-rotation coupling constant $c_{K0}$ can be written, approximately and succinctly, as 
\begin{equation}
\frac{1}{T_1} \propto \left(\frac{c_{K0} T}{h}\right)^2.
\end{equation}
(The full harmonic SRRS model is used for all fitting routines, which gives higher-order corrections to the overall temperature dependence---see Refs.~\onlinecite{Fitzgerald_1999,Patton_2007, Limes_2015} for further details.)
The temperature dependence of our data in the range 77~K--150~K has a shape similar to that predicted from the SRRS theory, suggesting that the relaxation mechanism is probably correct, but that a better understanding of the associated interaction strength might be needed. 
In their work on the SRRS theory, Happer and co-workers~\cite{Fitzgerald_1999} estimate the spin-rotation coupling constant to be $c_{K0}/h = -27$ Hz, via its proportionality to the known chemical-shift difference between gas and solid xenon \cite{Ramsey_1950}.
The SRRS calculation with $c_{K0}/h = -27$ Hz is shown as a solid black line in Fig.~\ref{fig:snow_ice_temp.pdf}.
This calculation discounts thermal vibrations when finding the paramagnetic and diamagnetic currents, which may affect the result. 
(We note that an independent {\em ab initio} calculation done in Ref.~\onlinecite{Fitzgerald_1999}, using a pseudopotential theory \cite{Wu_1985}, is $c_K/h = -14.1 |f|^2$ Hz, where $f$ is an introduced factor that accounts for any additional wave function overlap that was potentially not taken into account by their pseudopotential treatment.)
We can simply adjust the SRRS theory by forcing a fit through the 77 K ice data, which yields $c_{K0}/h = -22.3 \pm 0.2$ Hz, but this slightly overestimates the relaxation for the higher temperature ice data, as shown by the dashed line in Fig.~\ref{fig:snow_ice_temp.pdf}. 
If the ice data are fit excluding the 77 K and 155~K measurements, a value of $c_{K0}/h = -21.2 \pm 0.7$ Hz is found.

More recently, another approach involving pairwise-additive approximation (PAA) quantum chemical calculations was taken by Hanni, {\em et al.}~\cite{Hanni_2009}.
Using clusters of  up to 12 xenon atoms, they calculated, separately from first principles, nuclear shielding tensors (from which the chemical shift was obtained), nuclear quadrupole coupling tensors, and spin-rotation tensors.
By considering higher-than-pair interactions in an effective PAA model consisting of twelve xenon atoms, they predicted a nuclear shielding tensor for a central $^{129}$Xe atom ($C_{5 \nu}$) that leads to a chemical shift of $\delta \approx 322$ ppm---this matches remarkably well with the observed experimental gas-solid chemical shift of $^{129}$Xe, $\delta \approx 320$ ppm \cite{Raftery_1992}  (this shift is also confirmed by our measurements).
This lends support for the accuracy of their calculated spin-rotation constant for a central $^{129}$Xe atom in a twelve xenon cluster, $c_{K0}/h \approx -16.4 \pm 2$ Hz.
Substituting this value of the spin-rotation constant into the SRRS theory gives a value of the solid xenon $T_1$ to be approximately 310 min at 77 K.
One avenue to pursue in future work would be to consider more mechanisms or processes contributing to the relaxation that yield a similar temperature dependence, e.g., anharmonic two-phonon processes~\cite{vanKranendonk_1967,vanKranendonk_1968}.

To explore the possible effects of oxygen diffusion into both snow and ice samples, separate measurements were done in which we purposely introduced air into the sample chamber at specific times after the formation of the solid. 
In the case of ice at 77 K, no effect on the relaxation was observed, indicating that any mechanism mediated by oxygen diffusion into the ice is negligible. 
(We note that such a mechanism should also lead to a corresponding deviation from monoexponential decay, which was not observed for ice at 77 K.)  
This result stands in contrast to the case in which oxygen is introduced  {\em prior} to the formation of the solid: earlier studies of thermally polarized solid xenon (in samples that we would likely have termed "ice") have shown that $T_1$ values are dramatically reduced when oxygen is allowed to co-condense with xenon \cite{Warren_1966,Warren_1967}.
Hence, we conclude that very little oxygen was present at any time during the formation of our xenon ice samples at 77 K. 

We were further able to rule out paramagnetic relaxation due to {\em any} impurity by forming samples with both higher and lower $^{129}$Xe abundance compared with the naturally abundant (26.4\% $^{129}$Xe) samples that generated the data in Fig. \ref{fig:snow_ice_temp.pdf}. 
The two enriched-spin samples had 86\% spin-1/2 $^{129}$Xe and 1.8\% spin-3/2 $^{131}$Xe and were accumulated in the flow-through apparatus. 
The one dilute-spin sample was a convection cell made by mixing $^{130}$Xe with natural xenon to yield 4.5\% $^{129}$Xe and 4.3\% $^{131}$Xe.  
Measurements of $T_1$ in ice in these three samples at 77 K (shown in Fig. 3) are reasonably consistent with one another and with the value measured for the naturally abundant samples.  
From Refs.~\onlinecite{Cates_1990} and \onlinecite{Abragam_1982}, a dilute paramagnetic impurity creates a spin-diffusion barrier $b$ that should cause nuclei outside of $b$ to relax as
\begin{equation}
\frac{1}{T_{1n}} \propto \frac{(\Delta H_n)^2 N_e}{N_K},
\end{equation}
where $N_K$ is the number density of $^{129}$Xe nuclei and $N_e$ is the number density of impurities.
The nuclear linewidth $\Delta H_n$ is related to $N_K$ in order of magnitude by $\Delta H_n \sim \gamma_K \hbar N_K$, and to the diffusion barrier as $\Delta H_n \sim \gamma_S \hbar / b^3$. 
If $N_e$ increases, the relative concentration of impurities with frequency shifted nuclei to unshifted nuclei increases and $T_{1n}$ decreases. 
If $N_K$ increases, the linewidth of the unshifted nuclei increases, and in turn decreases the diffusion barrier $b$, allowing more shifted nuclei to contribute to relaxation and again leading to a decrease in $T_{1n}$.  
As $1/T_{1n} \propto N_K$, and as $N_K$ differs by as much as a factor of 20 across our ice data, this result reconfirms the absence of bulk relaxation mechanisms for $^{129}$Xe in solid xenon that are dependent on isotopic fraction, and demonstrates that the relaxation due to impurities in our ice samples is negligible. 
We note that the lack of dependence on $N_K$ also argues against any significant mechanism mediated by spin diffusion in the solid, for which we would expect both an $N_K$-dependence and non-monoexponential relaxation behavior.

\begin{figure}
\includegraphics{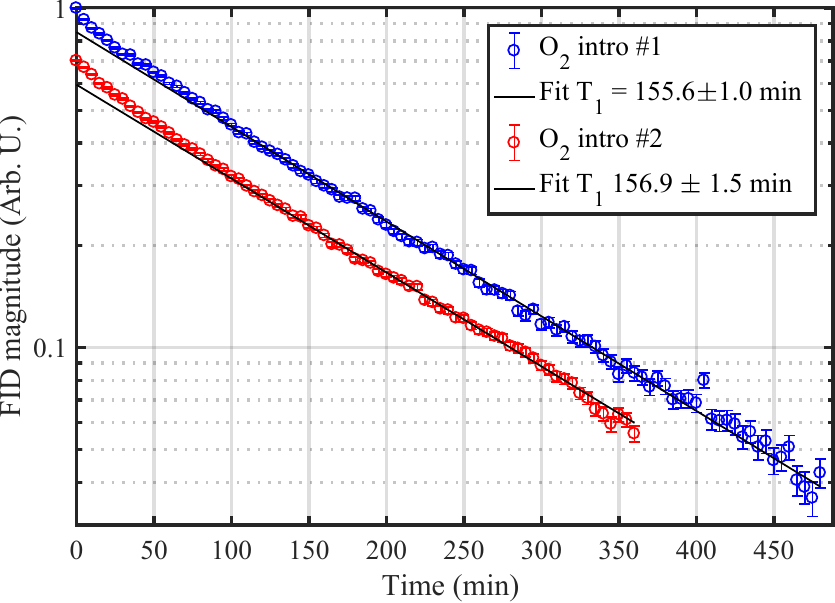}
\caption{ Snow $^{129}$Xe relaxation data taken at 77 K after introducing oxygen into the sample-chamber area are shown.  The early-time data points ($t < 100$ min) are masked in the fit and $T_1$ values are given for each late-time fit. The faster early-time rate is due to the destruction of surface polarization by adsorbed oxygen. Further relaxation is limited by $^{129}$Xe spin-diffusion to these surfaces.  }
\label{fig:post_formation_oxygen_doped_snow}
\end{figure}
We briefly note some potentially interesting observations concerning the snow data.
First, snow data taken above 100 K exhibited multi-exponential behavior, with, counter-intuitively, time-dependent $T_1$ that {\em decreased} with time at certain temperatures---the cause of this is not understood.
Second, as stated above, the possibility of oxygen as the cause of the ice/snow $T_1$ difference at 77 K prompted us to introduce air into our sample chamber after the formation of naturally abundant snow and ice. 
By doing so, an early-time increase of the relaxation rate in snow samples was found; the results of two such measurements are shown in Fig.~\ref{fig:post_formation_oxygen_doped_snow}.
The initial faster relaxation rate is governed by the adsorption/desorption rate and strength of the O$_2$-$^{129}$Xe interaction on the surface. 
 The relaxation  slows to a late-time rate, suggesting that the effect saturates when all of the polarization near crystallite surfaces has been destroyed. 
 The small increase in this late-time $T_1$ over that of the average snow  $T_1$ may indicate the surface oxygen creates a more well defined spin-diffusion boundary; essentially the oxygen may shift the frequencies of nearby surface $^{129}$Xe enough that the bulk $^{129}$Xe spins are less effective at diffusing to the exterior.
The oxygen would essentially ``protect'' the interior $^{129}$Xe from relaxing.
In any case, oxygen adsorbed onto the crystallite surface is clearly sufficient for a fast-relaxing boundary, but we have not determined if this process is the cause of the snow-ice discrepancy, or whether other surface-related physics might be at play.
Enriched snow runs were done that also showed an increased initial relaxation rate, but the overall picture remains ambiguous.
More experiments are required to make a consistent and complete description of our snow data and the precise nature of the difference between our obtained snow and ice $T_1$ data. 

\section{Conclusion}

The $^{129}$Xe ice $T_1$ data presented herein are the most reproducible to date, and in the case of the frozen liquid ice, also represent the longest T1 times measured at 77~K and above.
We show formation-method dependent $^{129}$Xe $T_1$ times, which we identify as snow and ice samples, are likely due to spin-diffusion to a fast-relaxing boundary, although the exact nature of the fast-relaxing boundary is not clear. 
We obtained ice $T_1$ times across 77-150 K that give a spin-rotation coupling strength roughly 20\% smaller than the previous publications that use the SRRS theory. 
There was no dependence of $T_1$ on the isotopic composition of the xenon, indicating no presence of bulk paramagnetic relaxation in ice.
Future experimental work should include a lower temperature range $T_1$ study (4-50 K) on ice and snow \cite{Lang_2002}, purposely introducing oxygen prior to and post solid formation in temperature dependent runs, and growing single-crystal solid Xe \cite{Sonnenblick_1982,Keyse_1985} to potentially reach an upper limit of $^{129}$Xe $T_1$ in efforts to better understand the spin-rotation coupling strength.

\begin{acknowledgments}
The authors thank M. Conradi, M. B.Walker, W. Happer, K. Kuzma, and G. Laicher for helpful discussions; and University of Utah glassblower K. Teaford for cell and sample chamber manufacture. This work was supported by the U.S. National Science Foundation, grant PHYS-0855482. M.E.L. thanks the Swigart Endowed Scholarship Fund at the University of Utah for support.
\end{acknowledgments}

%

\end{document}